\documentclass[english]{IEEEtran}
\usepackage[T1]{fontenc}
\usepackage{url}
\usepackage{amstext}
\usepackage{graphicx}

\makeatletter

\newcommand{\lyxdot}{.}

\usepackage{bm}

\makeatother

\usepackage{babel}
\begin{document}
Disclaimer: This work has been published in IEEE Transactions on Ultrasonics,
Ferroelectrics, and Frequency Control Vol. 61, No. 7, pp 1063-1074,
July 2014.

\url{http://dx.doi.org/10.1109/TUFFC.2014.3007}

Copyright with IEEE. Personal use of this material is permitted. However,
permission to reprint/republish this material for advertising or promotional
purposes or for creating new collective works for resale or redistribution
to servers or lists, or to reuse any copyrighted component of this
work in other works must be obtained from the IEEE. This material
is presented to ensure timely dissemination of scholarly and technical
work. Copyright and all rights therein are retained by authors or
by other copyright holders. All persons copying this information are
expected to adhere to the terms and constraints invoked by each author\textquoteright{}s
copyright. In most cases, these works may not be reposted without
the explicit permission of the copyright holder. For more details,
see the IEEE Copyright Policy

\clearpage

\title{Determination of Doping and Temperature Dependent Elastic Constants
of Degenerately Doped Silicon from MEMS Resonators}

\author{Antti Jaakkola, Mika Prunnila, Tuomas Pensala, James Dekker and Panu
Pekko\\
VTT Technical Research Centre of Finland, Espoo, Finland}
\maketitle
\begin{abstract}
Elastic constants $c_{11}$, $c_{12}$ and $c_{44}$ of degenerately
doped silicon are studied experimentally as a function of the doping
level and temperature. First and second order temperature coefficients
of the elastic constants are extracted from measured resonance frequencies
of a set of MEMS resonators fabricated on seven different wafers doped
with phosphorus (carrier concentrations $4.\text{1}\,,4.7\,\mbox{and\,}7.5\times10^{19}\mbox{cm }^{-3}$),
arsenic ($1.7\,\mbox{and\,}2.5\times10^{19}\mbox{cm }^{-3}$), and
boron $(0.6\,\mbox{and\,}3\times10^{19}\mbox{cm }^{-3}$), respectively.
Measurements cover a temperature range from $-40^{\circ}\mbox{C }$
to $+85{}^{\circ}\mbox{C }.$

It is found that that the linear temperature coefficient of the shear
elastic parameter $c_{11}-c_{12}$ is zero at n-type doping level
of $n\sim2\times10^{19}\mbox{cm }^{-3}$, and that it increases to
over $40\,\mbox{ppm/K}$ with increasing doping. This observation
implies that the frequency of many types of resonance modes, including
extensional bulk modes and flexural modes, can be temperature compensated
to first order. The second order temperature coefficient of $c_{11}-c_{12}$
is found to decrease by 40\% in magnitude when n-type doping is increased
from 4.1 to 7.5$\times10^{19}\mbox{cm }^{-3}$.

Results of this study enable calculation of the frequency drift of
an arbitrary silicon resonator design with an accuracy of $\pm25$~ppm
over $T=-40\ldots85^{\circ}\mbox{C}$ at the doping levels covered
in this work. Absolute frequency can be estimated with an accuracy
of $\pm1000$~ppm. 
\end{abstract}

\section{Introduction}

Single-crystal silicon MEMS resonators are challenging quartz devices
in timing and frequency control applications. The main disadvantage
of silicon resonators is their high frequency drift of about $-30\,\mbox{ppm/K}$,
which needs to be compensated to make a stable reference. Heavy doping
of silicon has recently been found as an attractive way to significantly
reduce this temperature dependency. Doping dependency of the elastic
constants of silicon can be explained as a free carrier effect. The
band structure of Si depends on strain and, therefore, the charge
carriers (introduced to the silicon crystal lattice with doping) redistribute
between different bands under strain \cite{keyes_electronic_1967,khan_temperature_1985}.
This leads to strain dependency of the carriers' free energy and introduces
doping dependent correction terms to the elastic constants. In n+
Si (p+ Si) the redistribution involves electrons (holes) that redistribute
between different conduction band minima (valence band maxima).

Doping based temperature compensation of silicon resonators started
with p-type doping \cite{samarao_passive_2010}, but n-type doping
soon appeared as a viable alternative \cite{hajjam_sub-100ppb/c_2010}.
Our work with bulk mode resonators has shown that n-type doping is
an effective and versatile way of tailoring the temperature behavior
of silicon resonators; we have demonstrated resonators with their
$f$~vs.~$T$ turnover point near room temperature, overcompensated
devices (+18~ppm/K) \cite{pensala_temperature_2011}, and shown that
n-type doping is applicable to virtually all resonance modes of practical
importance \cite{jaakkola_temperature_2012}. Recently, resonators
made of strongly n-type doped epitaxially grown silicon \cite{ng_localized_2013}
have been reported. 

The main contribution to the temperature dependent frequency drift
of a resonator comes from the elastic constants of the resonator material.
Thus, to optimize the thermal stability of a silicon MEMS resonator,
a designer needs to know the temperature behavior of the elastic parameters
of silicon; in particular the first and second order thermal derivatives
of the elastic constants are of interest. However, experimental data
of the temperature dependency of the elastic parameters of heavily
doped silicon is limited; most usable results of n-type doped silicon
have been published by Hall \cite{hall_electronic_1967} for carrier
concentration of $2\times10^{19}\mbox{cm }^{-3}$. 

In this work, silicon elastic constants $c_{11}$, $c_{12}$ and $c_{44}$
are studied experimentally as a function of doping level and temperature.
First and second order temperature coefficients of the elastic constants
are extracted from the resonance frequencies of a set of MEMS resonators
fabricated on seven different wafers with varied doping. In Section
\ref{sec:Methods}, the analysis method for extracting the unknown
elastic parameters from the measured $f$ vs. $T$ curves of the resonators
is introduced. The fabrication of the devices and the measurements
are covered in Section \ref{sub:experimental}. Results are presented
in \ref{sec:Results}. Implications of the results are discussed in
Section \ref{sec:Discussion}, concentrating on the aspects important
for temperature compensation of MEMS resonators. Reliability of the
elastic parameter extraction procedure is assessed, and, this provides
a way to estimate how accurately the frequency and its thermal drift
of an arbitrary resonator design can be calculated. An error analysis
of the extracted elastic parameters is presented, and, additionally,
MEMS resonator manufacturability aspects are covered.

\section{Methods\label{sec:Methods}}

\subsection{\label{sub:Extraction-of-the}Extraction of elastic constants from
resonance frequencies}

\begin{figure}[h]
\begin{centering}
\includegraphics[bb=0bp 0bp 600bp 640bp,clip,width=1\columnwidth]{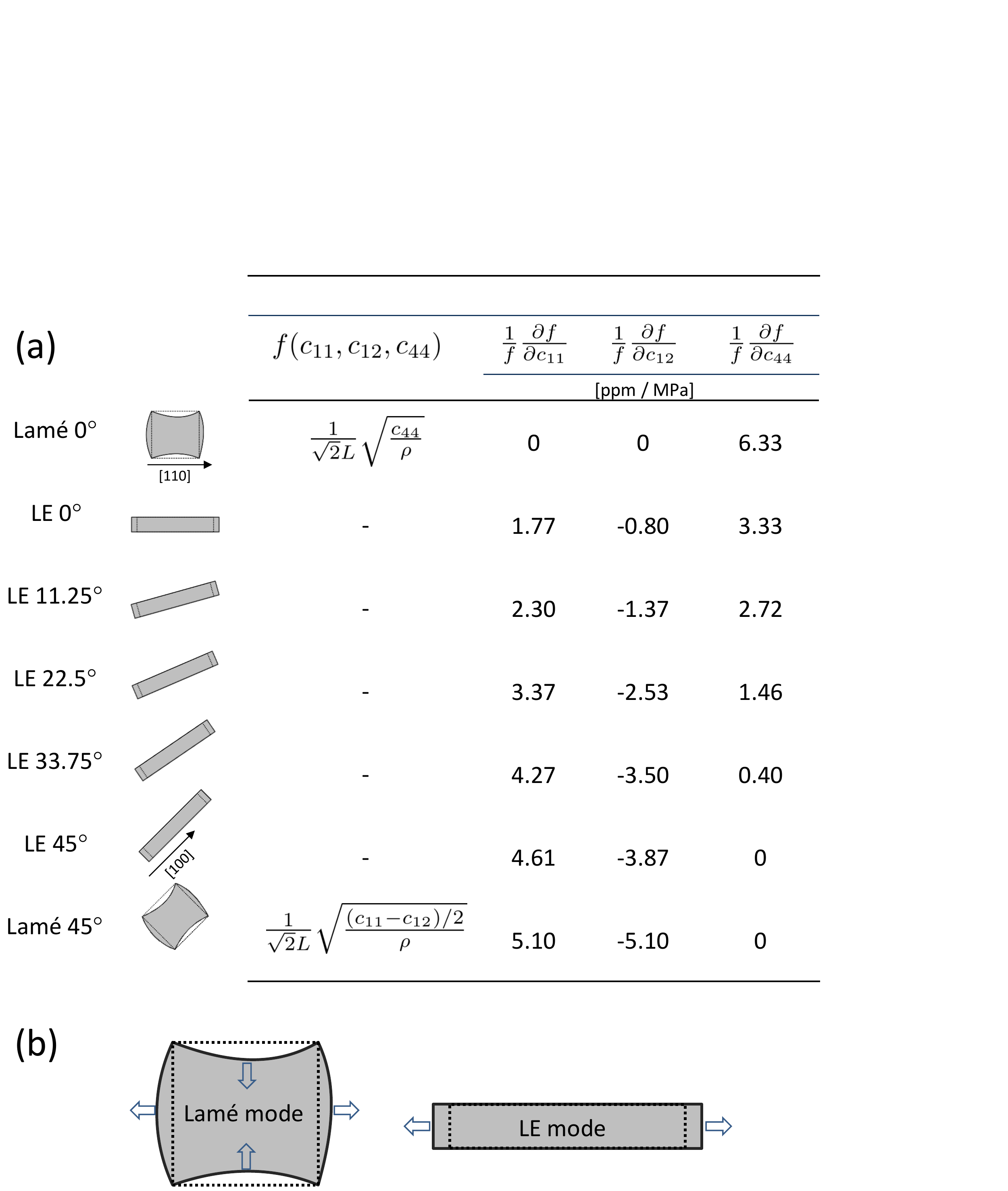}
\par\end{centering}

\caption{(a) Determination of the elastic constants $c_{11}$, $c_{12}$ and
$c_{44}$ is based on seven resonance modes, whose frequencies have
different dependencies on the $c_{ij}$ parameters. Alignment of the
resonators is varied from {[}110{]}\ to {[}100{]}. The table contains
the analytical formulas for the resonance frequency $f(c_{11},c_{12},c_{44})$
-- which exist only for the two Lamé modes -- and the sensitivities
$1/f\times\partial f/\partial c_{ij}$ for each mode. These exemplary
sensitivities have been calculated at a linearization point of $(c_{11},c_{12},c_{44})=(163,\,65,\,79)\,\mbox{GPa}$
using the finite element approach outlined in Section \ref{sub:Numerical-modelling}
(zero angular alignment error and device layer thickness $15\mbox{\,\ensuremath{\mu}m}$
of has been assumed). The listed numbers only illustrate the character
of the variation of the sensitivies within the set of modes. For individual
wafers, the sensitivities differ due to different linearization points,
device layer thicknesses and angular misalignments, respectively.
\label{fig:seven_modes} (b) Illustration of the mode shapes of the
Lamé/LE resonances.}
\end{figure}

The frequency of an acoustic resonator is given by
\begin{equation}
f=1/L\times\sqrt{c/\rho},\label{eq:frquency-general}
\end{equation}
where $\rho$, $c$ and $L$ are the resonator material density, characteristic
stiffness and characteristic length, respectively. The characteristic
stiffness depends on the elastic constants $c_{11}$, $c_{12}$ and
$c_{44}$ that can be solved from a set of measured resonance frequencies
of different resonance modes when their functional dependency on constants
$c_{ij}$ varies among the modes, and when there are three or more
modes within the set. In our case, the set of two Lamé mode resonators
and five length extensional (LE) modes in different orientations fulfill
these conditions. The test resonator set, and their exemplary sensitivities
on the $c_{ij}$ parameters are illustrated Fig. \ref{fig:seven_modes},
and micrographs of the two types of resonators are shown in Fig. \ref{fig:u_graph_and_mesh}(a).
Additional constraints that lead to the selection of this particular
set of devices were: 1) the resonators had to be actuated electrostatically
over vertical coupling gaps, 2) air damping needed to low enough to
allow detection of resonances in atmospheric pressure, 3) the resonators
had to be relatively large in lateral dimensions to minimize effects
from processing inaccuracies, 4) the resonance frequencies and their
sensitivities on $c_{ij}$ should be insensitive to device thickness
variations (see error $E_{5}$ in Section \ref{sub:Error-analysis}),
and 5) the number of different resonator types had to be relatively
large in comparison with the three unknowns $c_{ij}$ to allow assessment
of the reliability of the results (Section \ref{sub:Reliability-discussion}).

The experimental data consists of measured resonance frequencies at
different temperatures for all seven resonance modes $f_{k}^{\mbox{exp}}(T)$~($k=1,\ldots,7$).
Let us denote the corresponding theoretical estimates containing the
$c_{ij}$ dependencies -- obtained through FEM modelling -- as $f_{k}^{\mbox{th}}(c_{11},c_{12},c_{44})$.
We use an approach of first matching $f_{k}^{\mbox{exp}}$ and $f_{k}^{\mbox{th}}$
at $T_{0}=20^{\circ}\mbox{C}$ by numerical minimization of 
\begin{equation}
g(c_{ij})=\sum_{k}[f_{k}^{\mbox{exp}}-f_{k}^{\mbox{th}}(c_{ij})]^{2}\label{eq:optim_for_cij0}
\end{equation}
to find elastic parameters $c_{ij}(T_{0})$, and then linearize (\ref{eq:frquency-general})
to obtain the relation
\begin{equation}
\frac{\delta f_{k}^{\mbox{exp}}(T)}{f_{k0}^{\mbox{exp}}}=\frac{1}{f_{k0}^{\mbox{th}}}\sum_{ij}\frac{\partial f_{k}^{\mbox{th}}}{\partial c_{ij}}\delta c_{ij}(T)+\frac{1}{2}\frac{\delta L}{L}(T).\label{eq:linearized_01}
\end{equation}
Here $\delta c_{ij}(T)$ are the unknown changes in elastic parameters,
$\delta f_{k}^{\mbox{exp}}(T)$ are the measured frequency differences,
while $f_{k0}^{\mbox{exp}}$ and $f_{k0}^{\mbox{th}}$ are shorthands
for $f_{k}^{\mbox{exp}}(T_{0})$ and $f_{k}^{\mbox{th}}(c_{ij}(T_{0}))$.
Sensitivities $\partial f_{k}^{\mbox{th}}/\partial c_{ij}$ are calculated
from the theoretical estimates. The last term accounts for thermal
expansion, and it has been obtained by employing the isotropic of
nature of length changes for silicon. We use a 3rd order expansion
for this term 
\begin{equation}
\frac{1}{2}\frac{\delta L}{L}(T)=(\alpha_{1}\Delta T+\alpha_{2}\Delta T^{2}+\alpha_{3}\Delta T^{3})/2,\label{eq:thermalexp}
\end{equation}
where values of $\alpha_{1}=2.84\times10^{-6}\,\mbox{K}^{-1}$, $\alpha_{2}=8.5\times10^{-9}\,\mbox{K}^{-2}$
and $\alpha_{3}=-32\times10^{-12}\,\mbox{K}^{-3}$ are assumed. These
expansion coefficients are based on the values reported for undoped
silicon in Ref. \cite{lyon_linear_1977}; thus, it is assumed that
thermal expansion is not affected by doping. The assumption is supported
by our measurements with mechanical dilatometry (see Section \ref{sub:Doping-independency-of}).

In matrix form (\ref{eq:linearized_01}) can be denoted as 
\begin{equation}
\delta\mathbf{f^{\mbox{exp}}}(T)=A\cdot\delta\mathbf{c}(T)+\mathbf{\beta}(T).\label{eq:dfrel_in_matrixform}
\end{equation}
where $\delta\mathbf{f^{\mbox{exp}}}$ contains the relative frequency
changes, and elements of the sensitivity matrix $A$ are defined as
\begin{equation}
a_{kn}=\frac{1}{f_{k0}^{\mbox{th}}}\frac{\partial f_{k}^{\mbox{th}}}{\partial c_{n}},\, n=11,12,44;\, k=1,...,7.\label{eq:amatrix}
\end{equation}
Sensitivity matrix elements are illustrated in Fig. \ref{fig:seven_modes}.
One should note that (\ref{eq:amatrix}) depends on the linearization
point at which it is evaluated.

The changes in elastic parameters $\delta\mathbf{c}(T)$ can be solved
as a least squares fit from (\ref{eq:dfrel_in_matrixform}): 
\begin{equation}
\delta\mathbf{c}(T)=(A^{T}A)^{-1}A^{T}[\delta\mathbf{f^{\mbox{exp}}}(T)-\mathbf{\beta}(T)].\label{eq:LSQ}
\end{equation}

\subsection{Numerical modelling \label{sub:Numerical-modelling}}

Numerical estimates of the modal frequencies $f_{k}^{\mbox{th}}(c_{11},c_{12,},c_{44})$
were calculated by finite element analysis with Comsol Multiphysics.
Resonance frequencies were obtained with modal analysis of full 3D
geometries of the devices including the anchoring regions, see illustration
of the finite element mesh in Fig \ref{fig:u_graph_and_mesh}(b).
Nominal thickness of 15/24~$\mu\mbox{m }$ of the wafers were used
in the calculations, see Table \ref{tab:wafers_table}. Maximum size
of the mesh elements was $20\times20\times4\,\mu\mbox{m}^{3}$.

Calculation was performed for parameters $c_{11}$, $c_{12}$ and
$c_{44}$ spanning ranges of $[160\ldots168\,\mbox{GPa}]$, $[63\ldots68\,\mbox{GPa}]$
and $[78\ldots80\,\mbox{GPa}]$, respectively. The ranges were discretized
to a grid of $5\times5\times5$ points. Values were stored in tables,
and later retrieved for evaluation of Eqs. \ref{eq:optim_for_cij0}
and \ref{eq:amatrix}. Cubic interpolation was used for evaluation
of $f_{k}^{\mbox{th}}(c_{11},c_{12,},c_{44})$ between grid points.
The discretization was verified to be dense enough for accurate evaluation
of the derivatives of (\ref{eq:amatrix}). Simulations took into account
different angular misalignments of the wafers.

\subsection{Measurement of angular misalignment\label{sub:Detection-of-angular}}

\begin{figure}[h]
\begin{centering}
\includegraphics[bb=0bp 10bp 260bp 220bp,clip,width=0.9\columnwidth]{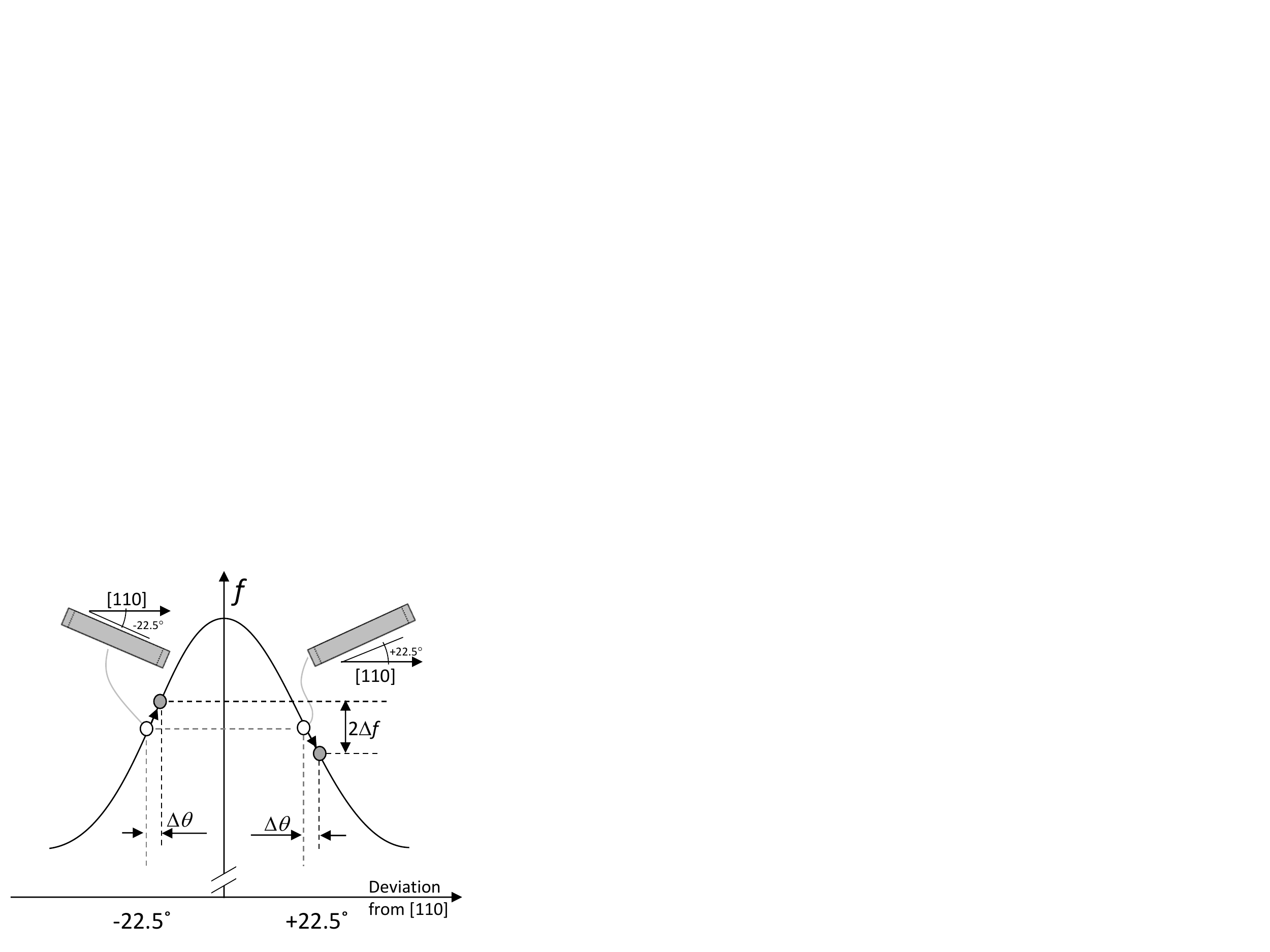}
\par\end{centering}

\caption{In-plane angular misalignment of the wafers can be deduced from the
difference of the resonance frequencies of two types of LE beam resonators,
which are designed at an angle of $\pm22.5^{\circ}$ from {[}110{]}
direction. In-plane rotation of the resonators shifts the resonance
frequencies up/down by $\pm460\,\mbox{ppm}/0.1\,\mbox{deg}$. \label{fig:aoffset_principle}}
\end{figure}

In practice, fabrication of the devices results in a small deviation
of the resonator orientation from the intended alignment with the
crystal axes, which can affect accuracy of the extraction of elastic
parameters. This deviation, or angular misalignment $\Delta\theta$,
was determined using the method illustrated in Fig. \ref{fig:aoffset_principle}.
Due to silicon anisotropy, the resonance frequency of a LE mode beam
resonator increases by \textasciitilde{}10\% when resonator alignment
is rotated from {[}100{]} to {[}110{]}. Between these directions,
i.e., at $\pm22.5^{\circ}$ from {[}110{]}, the resonance frequency
is most sensitive to angular misalignment with $\Delta f/\Delta\theta\sim\pm460\,\mbox{ppm}/0.1\,\mbox{deg.}$
Copies of two LE resonators identical in dimensions, but oriented
$45{}^{\circ}$ to each other, both at the most sensitive orientation
of $\pm22.5{}^{\circ}$were included on the wafers. The angular misalignment
could be deduced from the up/down frequency shifts $\pm\Delta f$
of these resonators.

\section{Experimental\label{sub:experimental}}

\begin{table}[h]
\caption{Details of the wafers for the silicon device layers. Carrier concentrations
were calculated from the resistivity specification using Ref. \cite{_standard_2000}.
Carrier concentration ranges are included as error bars in Fig. \ref{fig:CIJ_matrix}.
Angular misalignment was measured using the method described in Section
\ref{sub:Detection-of-angular}. \label{tab:wafers_table}}

\centering{}\includegraphics[width=1\columnwidth]{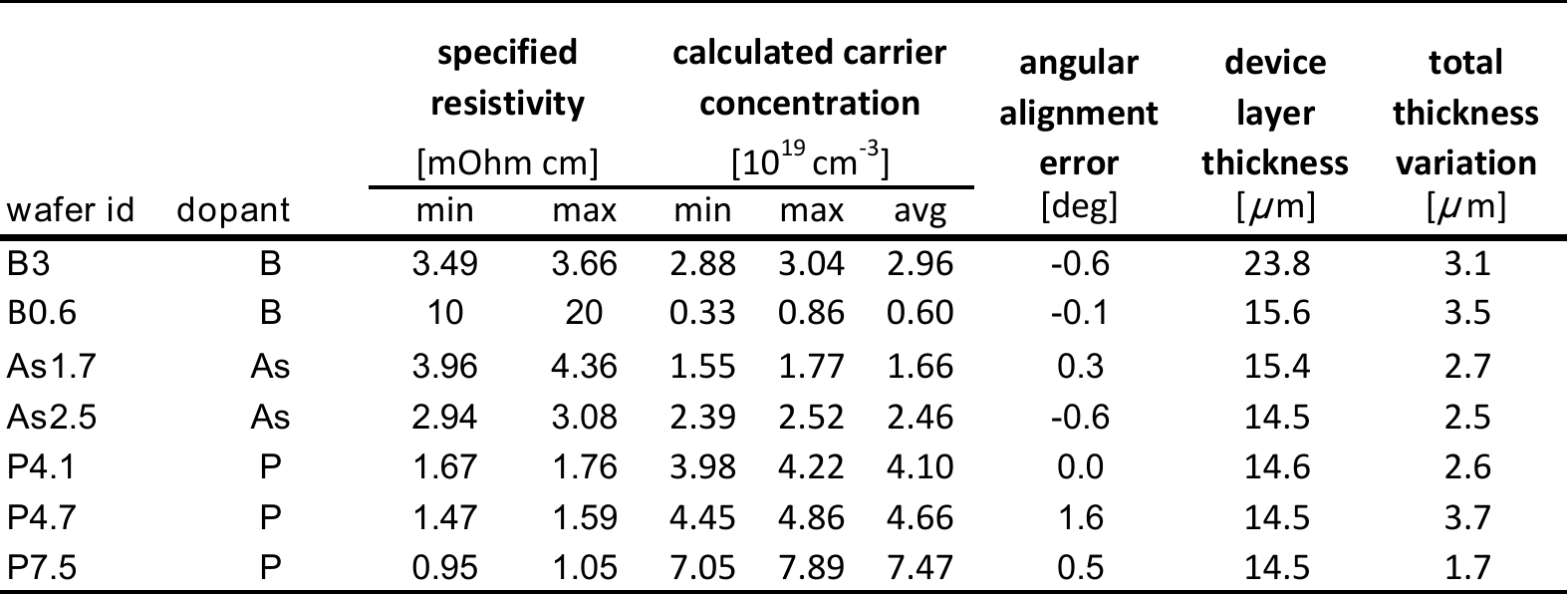}
\end{table}
\begin{figure}[h]
\begin{centering}
\includegraphics[bb=0bp 20bp 720bp 520bp,clip,width=1\columnwidth]{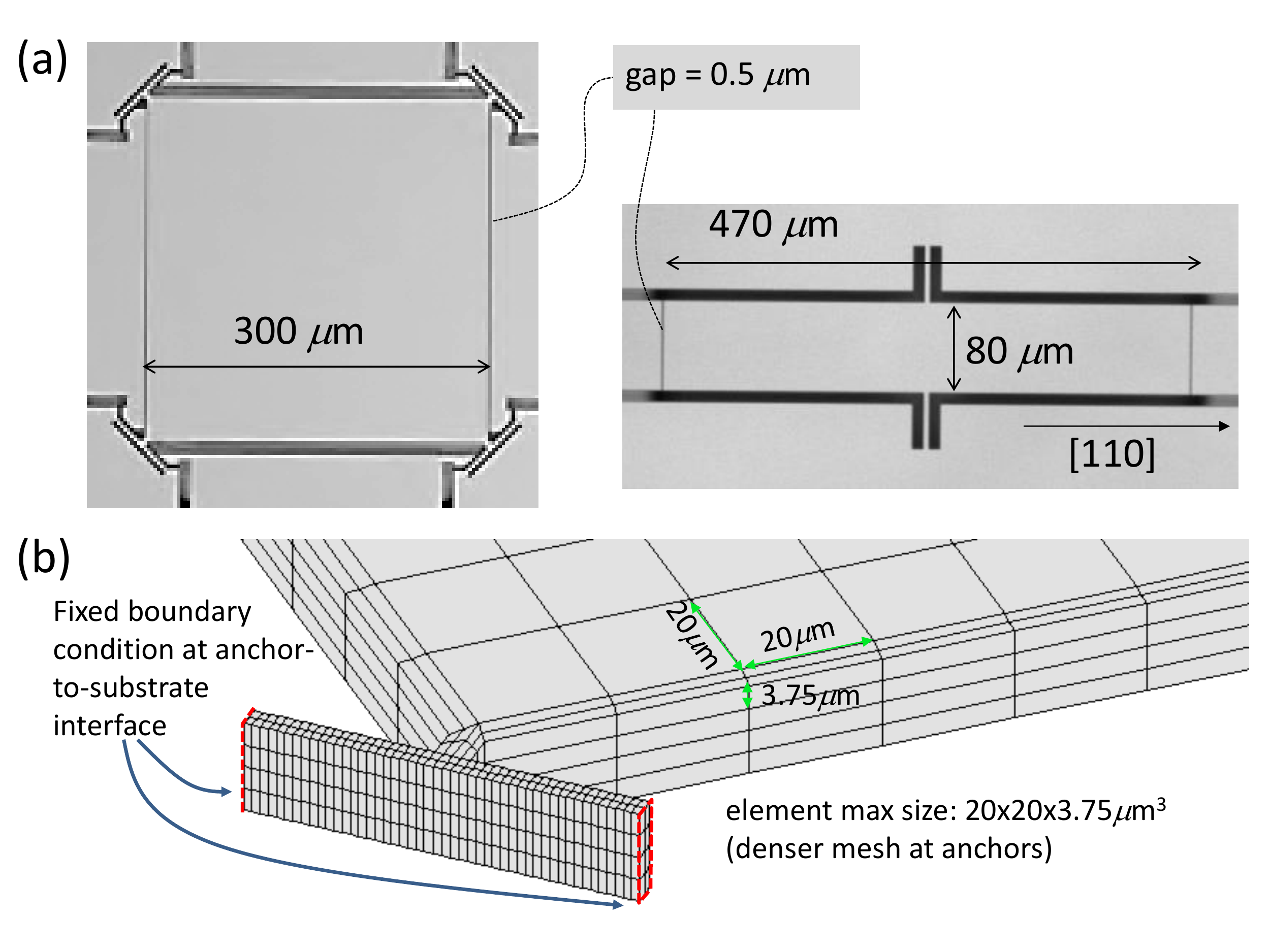}
\par\end{centering}

\caption{(a) Micrographs of Lamé and LE mode resonators. (b) Illustration of
the finite element mesh at one corner of the Lamé mode resonator.
Meshing was done in similar fashion for the LE mode resonator models.
\label{fig:u_graph_and_mesh}}
\end{figure}
The resonators (Fig. \ref{fig:u_graph_and_mesh}(a)) were fabricated
on seven different 150-mm C-SOI wafers (Silicon-On-Insulator wafers
with pre-etched cavities \cite{luoto_mems_2007}) manufactured in
co-operation with Okmetic Oyj. The handle wafers with DRIE etched
cavities were thermally oxidized before they were fusion bonded to
the device wafers. Device layers were fabricated from 100-oriented
wafers grown with the Czochralski method. The resonator fabrication
process started with the C-SOI wafers with circular cavities of a
diameter of $500\,\mbox{\ensuremath{\mu}m}$ for each resonator. The
process flow consisted of two lithographic layers: 1) Al contact metallization
and patterning 2) DRIE release etch producing vertical gaps of minimum
nominal width of $0.5\,\mu\mbox{m}$. 

Dopant (B, P, and As) concentrations of the wafers for the silicon
device layer were varied according to Table \ref{tab:wafers_table}.
Carrier concentration range for each wafer was calculated from the
specified resistivity range using conversion method of Ref. \cite{_standard_2000}.
Device layer nominal thickness was 15~$\mu\mbox{m}$ (24~$\mu\mbox{m}$
for wafer B3), and the manufacturer specified C-SOI stack total thickness
variation (including the handle wafer) was within $\pm2\,\mu\mbox{m}$
for all wafers.

As the wafers featured pre-etched cavities, it was possible to fabricate
monolithic resonators without a grid of release etch holes within
the devices, and thus the elastic properties of the resonators could
be accurately modelled. In our previous studies \cite{pensala_temperature_2011},
existence of release etch holes was a source of uncertainty for the
determination of the elastic constants. 

\begin{figure}[h]
\begin{centering}
\includegraphics[bb=0bp 10bp 410bp 330bp,clip,width=1\columnwidth]{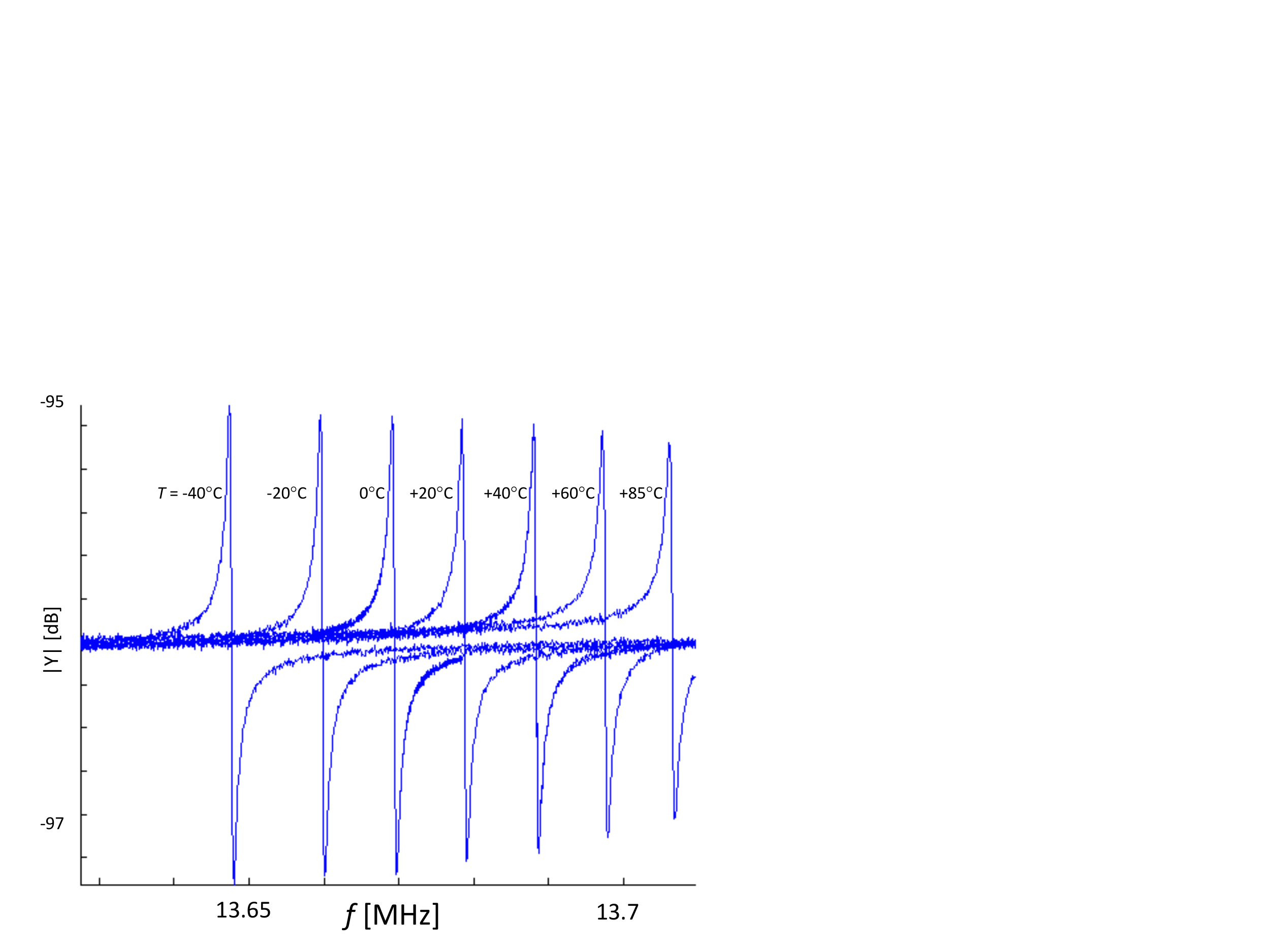}
\par\end{centering}

\caption{Measured admittance traces of the Lame-$45{}^{\circ}$ resonators
on wafer P7.5 measured at different temperatures. \label{fig:example_traces}}
\end{figure}
The resonators were measured on wafer level in atmospheric pressure
on a Cascade Summit probe station using a HP 4294A impedance analyzer.
A two-needle probe card was used for the measurements, and an open-short-load
calibration was performed at $T=40^{\circ}\mbox{C}$ in the beginning
of the measurement. Measurements were done in the four-terminal pair
configuration with four 2-m BNC cables, and the two end connections
to the probe needles were $\sim15\,\mbox{cm }$ long. A DC bias voltage
of 40 V was applied between the resonator and the electrodes for electromechanical
coupling. Effect on the resonance frequencies from the DC bias was
negligible because of the relatively wide coupling gaps and the high
mechanical spring constant of the resonance modes. The excitation
AC voltage level was set to 1 V to maximize signal-to-noise ratio.
The resonators still operated at their linear regime due to the weak
electromechanical coupling. Measured devices were located near the
wafer center. Quality factors of $Q\sim10,000$ were measured for
all resonance modes, and the resonance frequencies were extracted
by fitting a BVD equivalent circuit to the the measured admittances.
Example traces are shown in Fig. \ref{fig:example_traces}. 

The wafer was held on a temperature controlled chuck, whose temperature
was varied from -40$^{\circ}$C to +85$^{\circ}$C with seven steps
(for wafer As1.7 the highest temperature was 80$^{\circ}$C). The
specified temperature accuracy of the system (Temptronic TP3200A)
including the temperature controller and the chuck was $\pm0.5^{\circ}\mbox{C}$.
A 15 minute stabilization period followed after each temperature change
before probing of the resonators was started; the chuck temperature
was well stabilized in less than 10 minutes for all temperature steps.
Clean dry air flow at a rate of 30~l/min was used for purging. The
effect from room temperature gas flow to resonator temperature was
found to be smaller than the specified uncertainty of $\pm0.5^{\circ}\mbox{C}$
by the following comparison: Lame-45$^{\circ}$ resonator $f$ vs.
$T$ curves on wafer B0.6 were compared to corresponding data from
a similar wafer that was wafer level encapsulated by a silicon/glass
wafer (encapsulation method is described in Ref. \cite{kaajakari_stability_2006}).
Encapsulated resonators can be assumed to be free from thermal gradients
caused by the gas flow, but it could potentially affect temperature
of resonators on uncapped wafers like B0.6. $f$ vs. $T$ curves of
resonators near the wafer center on these two wafers were found to
overlap with each other within $15$~ppm, implying that device temperatures
were within $\sim0.5{}^{\circ}$C with each other (assuming identical
temperature coefficients for the resonators on both wafers (see Table
\ref{tab:spiders-table}).

\section{Results\label{sec:Results}}

\begin{figure}
\begin{centering}
\includegraphics[width=1\columnwidth]{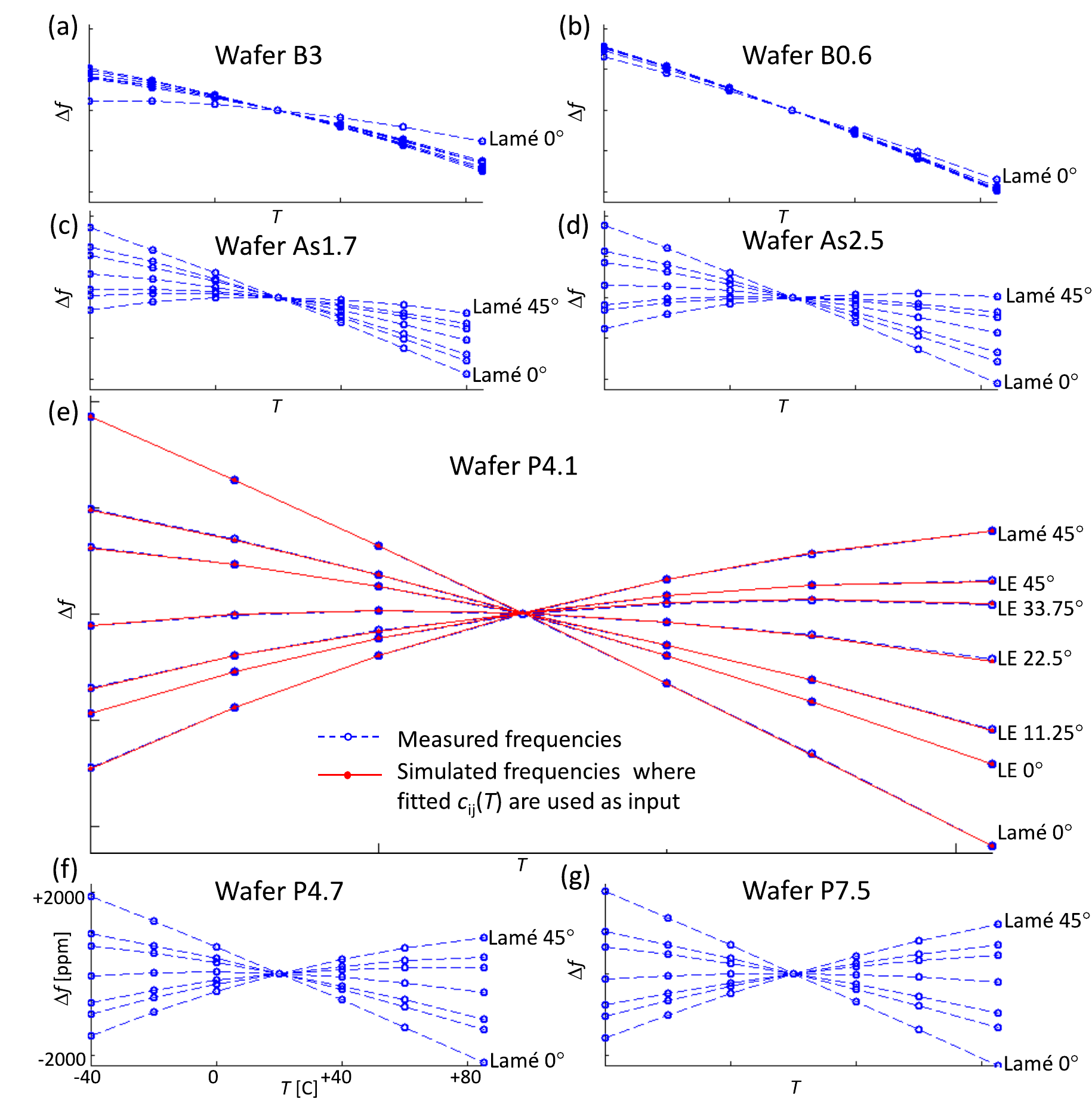}
\par\end{centering}

\caption{(a)--(g): Measured $f$ vs. $T$ data of all resonance modes on all
wafers is shown with blue open circles. Dashed blue lines are second
order polynomial fits to the data; fit coefficients are collected
in Table \ref{tab:spiders-table}. All plots have the similar scaling
of axes. (e): Experimental data from wafer P4.1 has been overlaid
with corresponding numerical estimates $f_{k}^{\mbox{th}}(T)$ which
use the fitted parameters $c_{ij}(T)$ as an input (red lines with
dots). The overlap is shown in more detail in Fig. \ref{fig:errrorfigs}(b).\label{fig:spiders}}
\end{figure}

Measured frequency vs. temperature curves are shown for all modes
on all wafers in Fig. \ref{fig:spiders}, and the related temperature
coefficients are collected in Table \ref{tab:spiders-table}.
\begin{table*}
\centering{}\ref{tab:spiders-table}\caption{Temperature coefficients of frequency ($f_{0}$, $a$ and $b$) of
the resonance modes of Fig. \ref{fig:spiders}. Fit was done to polynomial
$f(T)=f_{0}[1+a(T-T_{0})+b(T-T_{0})^{2}],$ which was centered at
$T_{0}=25^{\circ}\mbox{C}$. The fits reproduced the $f$ vs. $T$
curves to within $\pm10\,\mbox{ppm}$ for all cases.\label{tab:spiders-table}}
\includegraphics[width=1\textwidth]{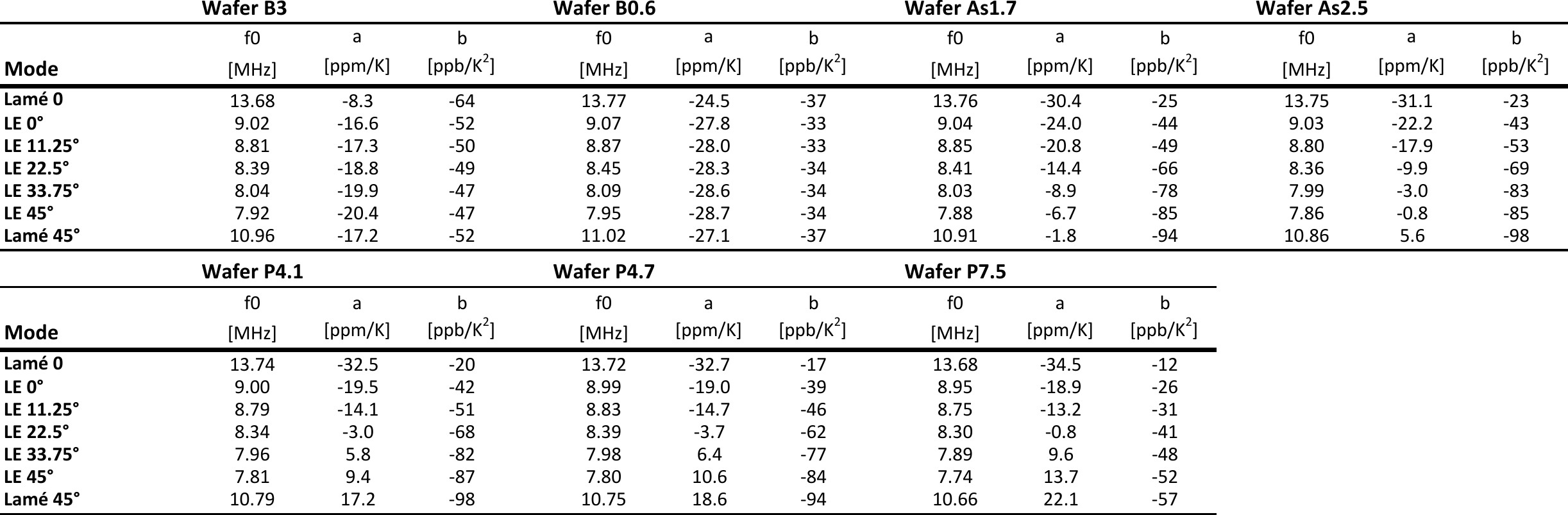}
\end{table*}
 On the weakest doped wafer B0.6 all $f$ vs. $T$ curves lie almost
on top of each other, and the linear temperature coefficients are
near -30~$\mbox{ppm\ensuremath{/}K}$. On wafer B3 , the slopes of
the curves are decreased in magnitude, and the biggest change is observed
for the Lamé-$0^{\circ}$ mode. On n-type doped wafers larger effects
are observed. The slope of the $f$ vs. $T$ curve of the Lamé-$45^{\circ}$
mode is gradually increased with increasing doping, and above $2\times10^{19}\mbox{\,\ cm}^{-3}$
the slopes are positive. Lame-$0{}^{\circ}$ mode is almost unaffected
by doping, and the $f$ vs. $T$ curves of the LE modes span the region
between the two Lamé modes.

\begin{figure}[h]
\begin{centering}
\includegraphics[bb=0bp 10bp 370bp 210bp,clip,width=1\columnwidth]{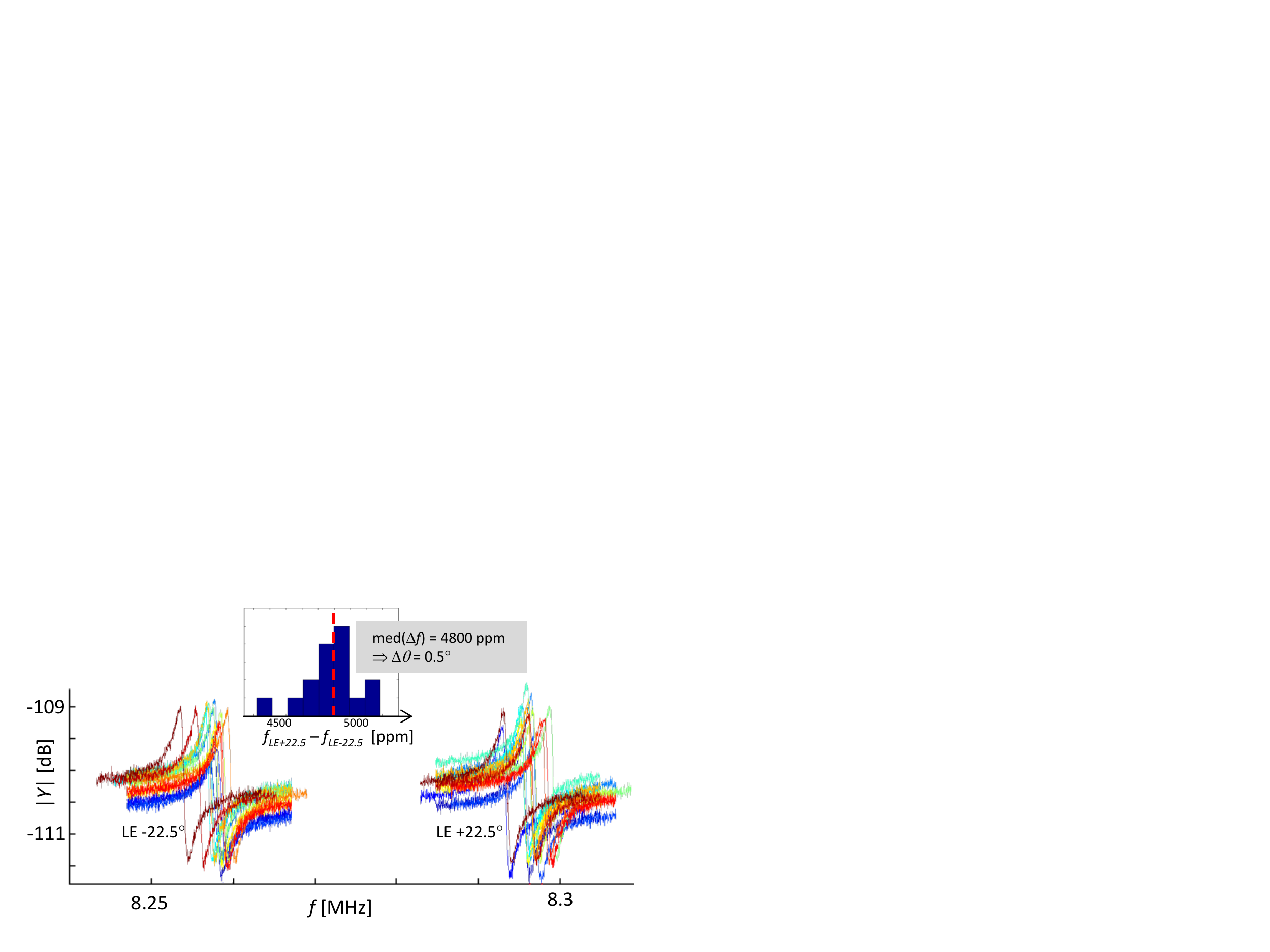}
\par\end{centering}

\caption{Admittance traces of 32 LE mode beam resonators (16 pairs) on wafer
P7.5 at $\pm22.5^{\circ}$ offset from the {[}110{]} direction. Closest
pairs (one resonator with $+22.5^{\circ}$ offset and the other with
$-22.5^{\circ}$, respectively) have been colored similarly, and all
pairs have different colors. Inset: Distribution of the frequency
differences between the closest pairs. Angular misalignment is calculated
from the median of this distribution (see Section \ref{sub:Detection-of-angular}).
\label{fig:aoffset_example}}
\end{figure}

Frequencies of >20 LE beam resonators were measured on each wafer
for determination of the angular misalignment as described in Section
\ref{sub:Detection-of-angular}. Results are tabulated in Table \ref{tab:wafers_table}.
Figure \ref{fig:aoffset_example} shows an example of the resonance
curves for wafer P7.5. 

\begin{figure}[h]
\begin{centering}
\includegraphics[bb=0bp 0bp 720bp 540bp,clip,width=1\columnwidth]{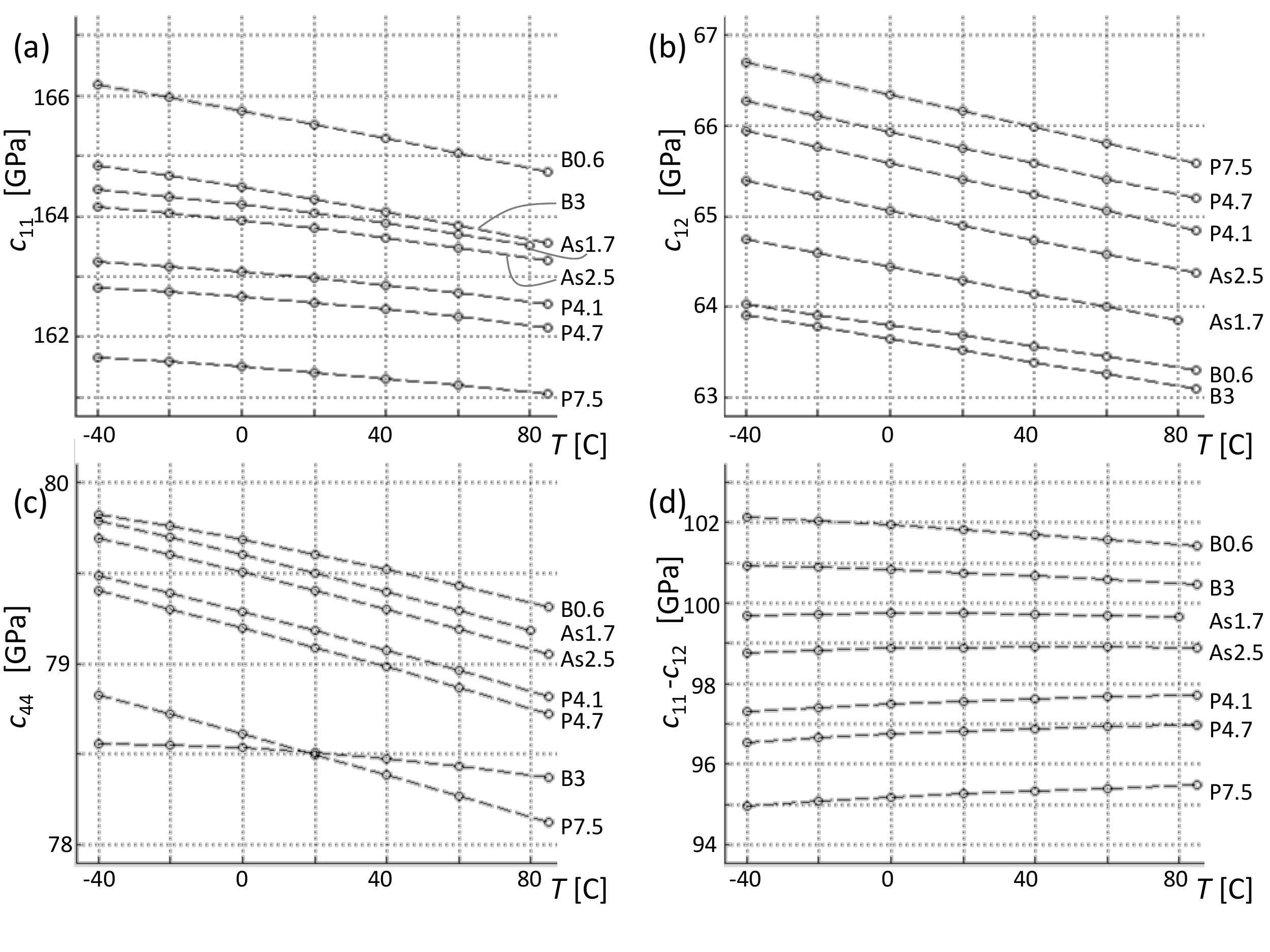}
\par\end{centering}

\caption{Elastic parameters $c_{11}$, $c_{12}$, $c_{44}$, and $c_{11}-c_{12}$
as a function of temperature and doping. The legends denote the dopant
element types and the doping level, see Table \ref{tab:wafers_table}.
Dashed lines are second order fits to the $c_{ij}$ vs. $T$ data,
and the fit coefficients are displayed in Fig. \ref{fig:CIJ_matrix}
and in Table \ref{tab:cij_results_table}. \label{fig:cij_curves}}
\end{figure}

\begin{figure*}
\begin{centering}
\includegraphics[width=1\textwidth]{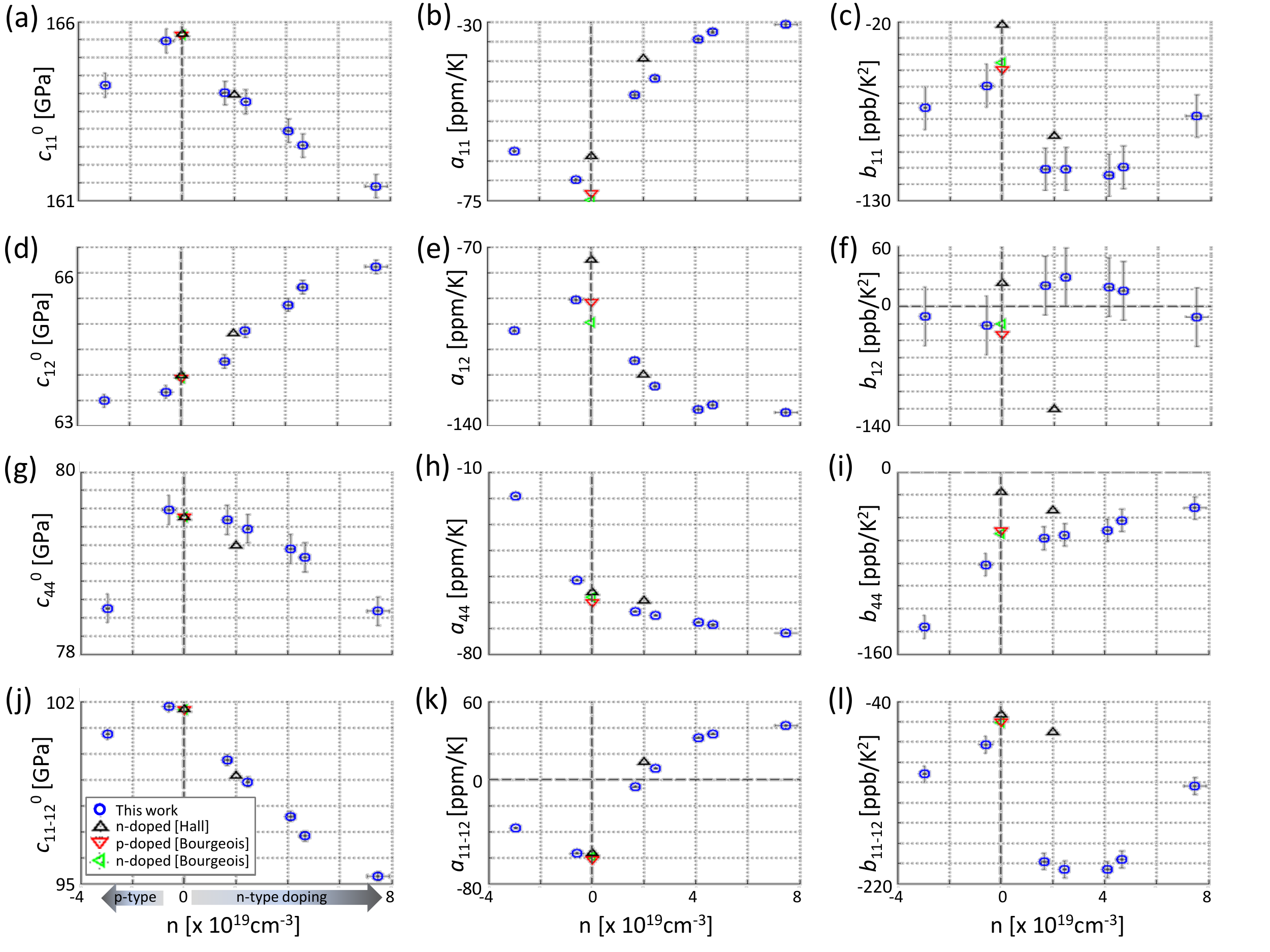}
\par\end{centering}

\caption{Temperature coefficients of the elastic parameters $c_{ij}$ as a
function of carrier concentration $n$. Data from p-type doped wafers
is represented with negative carrier concentrations. First, second
and third column represent the constant terms $c_{ij}^{0}$, linear
coefficients ($a_{ij}$), and second-order coeffients (\textbf{$b_{ij}$})
at $T=25^{\circ}C$, respectively, see (\ref{eq:polynomial}). $c_{11-12}^{0}$,
$a_{11-12}$ and $b_{11-12}$ are shorthands for the coefficients
of $c_{11}-c_{12}$. Open blue circles are the experimentally determined
values of this work. Numerical values are given in Table \ref{tab:cij_results_table}.
Values reported in \cite{bourgeois_design_1997} are shown as red
triangles pointing down (weak p-type doping) and as green triangles
pointing left (weak n-type doping). Data reported by Hall in Ref.
\cite{hall_electronic_1967} was used for calculating data points
shown as black triangles pointing up. Horizontal error bars indicate
the carrier concentration ranges calculated from the resistivity specification
for each wafer (see Table \ref{tab:wafers_table}). Vertical error
bars are based on the error analysis of Section \ref{sub:Error-analysis}.
\label{fig:CIJ_matrix}}
\end{figure*}

The elastic parameters $c_{ij}(T)$ were extracted from the measured
frequency data using the least squares method of Section \ref{sub:Extraction-of-the}.
Results are shown in Fig. \ref{fig:cij_curves}. The magnitude of
the elastic constants is observed to decrease upon increased doping,
except for the $c_{12}$ elastic constant which gets larger with increasing
n-type doping. 
For closer investigation of the thermal dependency of the $c_{ij}(T)$
curves, second-order polynomials centered at $T_{0}=25^{\circ}\mbox{C}$
were fitted to the elastic parameter data as
\begin{equation}
c_{ij}(T)=c_{ij}^{0}[1+a_{ij}(T-T_{0})+b_{ij}(T-T_{0})^{2}],\label{eq:polynomial}
\end{equation}
where $a_{ij}$ and $b_{ij}$ are the first-order and second-order
temperature coefficients, respectively, and $c_{ij}^{0}$ is the constant
term%
\footnote{One should notice that the least squares method of Section \ref{sub:Extraction-of-the}
uses $T=20^{\circ}\mbox{C}$ as the linearization point, since it
was one of the measurement points. However, expansions of (\ref{eq:polynomial})
and that of Table \ref{tab:spiders-table} are customarily centered
at $25^{\circ}\mbox{C}.$%
}. A second-order expansion of $c_{ij}(T)$ was found to be valid to
within $\pm20\,\mbox{ppm}$ for all $c_{ij}(T,n)$. The results are
collected in Figs. \ref{fig:CIJ_matrix}(a)--(l) and in Table \ref{tab:cij_results_table}.
One should note that, in Fig. \ref{fig:CIJ_matrix}, we have chosen
to accommodate data points from both n- and p-type doped wafers within
same axes by representing p/n type doping with negative/positive carrier
concentrations. Observations are discussed in the following section.

\begin{table*}
\caption{Values of temperature coefficients of the elastic parameters $c_{ij}$.
Corresponding data points are plotted in Fig. \ref{fig:CIJ_matrix}.
Confidence intervals $\Delta c_{ij}^{0}$, $\Delta a_{ij}$ and $\Delta b_{ij}$
are based on the error analysis of Section \ref{sub:Error-analysis}.
\label{tab:cij_results_table}}

\centering{}\includegraphics[width=1\textwidth]{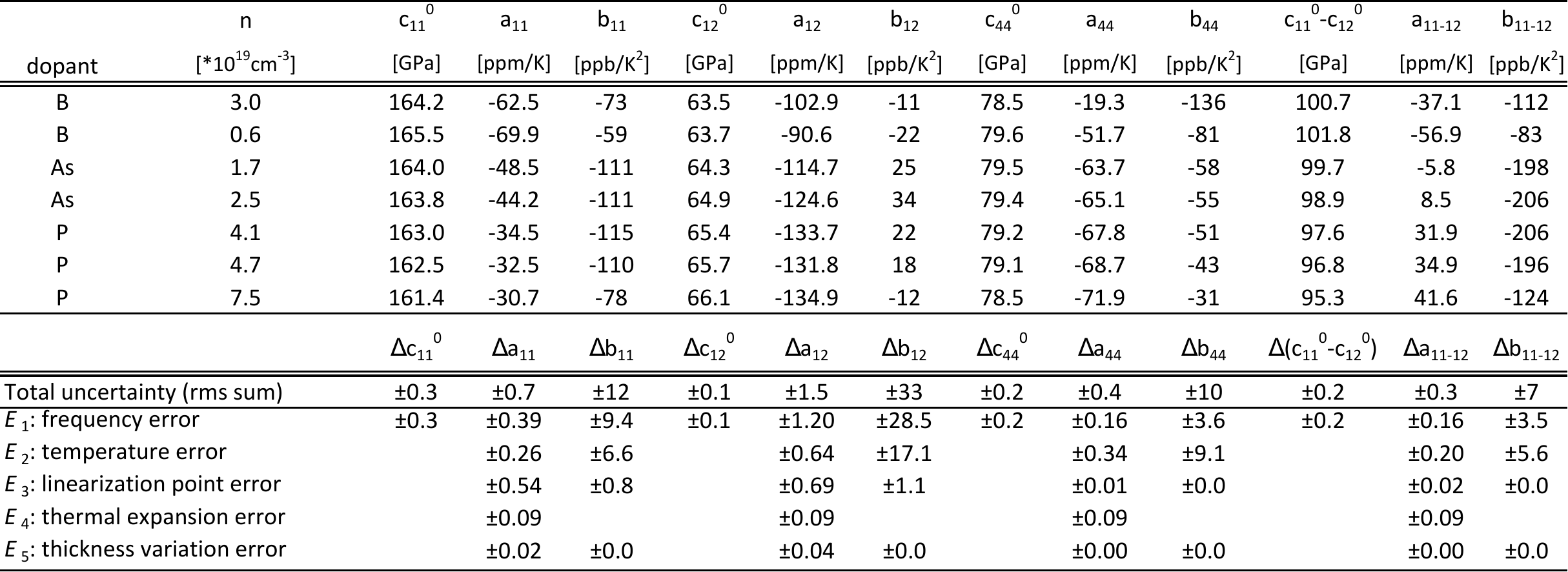}
\end{table*}


\section{Discussion \label{sec:Discussion}}

\subsection{Comparison to literature}

Temperature coefficients measured in this work are compared to previously
reported values in Fig. \ref{fig:CIJ_matrix}. Values for relatively
weakly n- or p-doped silicon, reported by Bourgeois et al. \cite{bourgeois_design_1997},
appear to be in satisfactory agreement with our data; the data points
near zero carrier concentration follow the trends observable from
our data points. Data by Hall \cite{hall_electronic_1967} differs
somewhat from our results, in particular for the second order temperature
coefficients. However, it should be noted that the temperature coefficients
for Hall's data are based on graphical extraction of the published
$c_{ij}(T)$ curves.

\subsection{Behavior of elastic coefficients with doping}

Figs. \ref{fig:CIJ_matrix}(a), (d), (g) and (j) show that the magnitude
of the elastic parameters, i.e., the constant terms $c_{ij}^{0}$,
are affected to within a few percent by increased doping over the
tested wafers. These changes need to be taken into account when dimensioning
resonator designs targeting a specific resonance frequency. While
the offsets have a negligible effect for temperature compensation
purposes, it has the potential to degrade the initial frequency accuracy
within a set of devices on a single wafer or within a batch of wafers.

In general, it is seen that arsenic (data points with $0<n<4\times10^{19}\,\mbox{cm}^{-3}$)
and phosphorus ($n>4\times10^{19}\,\mbox{cm}^{-3}$) as dopants do
not stand out from the plots as separate groups, which supports the
view of the effects being of mainly electronic origin\cite{keyes_electronic_1967}. 

The effects on silicon elastic properties from n-type doping are
best observed in the shear elastic constant $c_{11}-c_{12}$ and,
in particular, in its temperature coefficients $a_{11-12}$ and $b_{11-12}$.
Fig. \ref{fig:CIJ_matrix}(k) shows that the linear temperature coefficient
$a_{11-12}$ crosses zero at approximately $n=2\times10^{19}\mbox{\,\ cm}^{-3}$.
This is the effect of most practical importance for temperature compensation
of MEMS applications, since many shear-type resonance modes are purely
dependent on the $c_{11}-c_{12}$ term, and hence the linear temperature
coefficient of frequency of such resonators can be brought to zero
at this doping level.  For example, the Lame $45^{\circ}$ mode of
Fig \ref{fig:seven_modes}(a) is a mode whose frequency depends solely
on $c_{11}-c_{12}$. When doping is further increased, $a_{11-12}$
reaches a level of over $+40\,\mbox{ppm/K}$. The effect appears to
saturate with increasing doping. A wide class of resonance modes,
such as torsional, flexural and extensional modes have an amount of
shear mode character, i.e., their frequency depends on $c_{11}-c_{12}$
with a large weight factor. Thus, their $f$ vs. $T$ curves are largely
determined by the behavior of the $c_{11}-c_{12}$ term. The fact
that $a_{11-12}$ attains relatively large positive values enables
first-order temperature compensation of such modes. These aspects
are discussed in more detail in Ref. \cite{jaakkola_temperature_2012}. 

Fig. \ref{fig:CIJ_matrix}(l) shows that the second order coefficient
$b_{11-12}$ is negative for all studied doping levels, with a maximum
deviation from zero of approximately $-200\,\mbox{ppb/K}^{2}$. This
would translate to a 250 ppm frequency deviation over a range of 100$^{\circ}$C.
 Importantly, one finds that the second order coefficient $b_{11-12}$
appears to approach zero when n-type doping level is above $n=4.1\times10^{19}\mbox{\,\ cm}^{-3}$.
This suggests a possibility of a flat or positive second order response
at high enough doping, motivating further investigation of n-type
doping beyond $10^{20}\,\mbox{cm}^{-3}$. 

The key effect to temperature compensation with p-type doping is
observable in Fig. \ref{fig:CIJ_matrix}(h). The linear temperature
coefficient $a_{44}$ approaches zero with increasing p-type dopant
concentration. However, zero level is not yet crossed with the highest
doping level of $3\times10^{19}\mbox{\,\ cm}^{-3}$. The second order
coefficient $b_{44}$ is seen to grow in magnitude with increased
p-type doping.  N-type doping is observed to have a relatively small
effect on coefficients $a_{44}$ and $b_{44}.$

\subsection{Reliability of elastic parameter extraction\label{sub:Reliability-discussion}}

Seven data points were used for the extraction of the three unknown
elastic parameters $c_{ij}(T)$ at each temperature. Hence, the reliability
of the method can be assessed by comparing the measured frequency
data to the corresponding numerical estimates obtained from FEM simulations
which use the solved parameters $c_{ij}(T)$ as an input. First, Fig.
\ref{fig:errrorfigs}(a) shows the correspondence of measured and
simulated resonance frequencies at $T_{0}=20^{\circ}\mbox{C}$, where
$c_{ij}(T_{0})$ has been obtained from a fit to (\ref{eq:optim_for_cij0}).
The difference is within $\pm1000\,\mbox{ppm.}$

Correspondingly, the quality of the least squares fit of (\ref{eq:LSQ})
can be judged from the overlap of the measured and simulated data,
exemplified in Fig. \ref{fig:spiders}(e). This is seen in closer
detail in Fig. \ref{fig:errrorfigs}(b), where the difference between
the measured relative frequency changes $\delta f_{k}^{\mbox{exp}}(T)$
and the corresponding theoretical estimates $\delta f_{k}^{\mbox{th}}(c_{ij}(T))$
has been plotted for all modes on all wafers. Maximum deviation between
the measured and simulated data points was below $25\,\mbox{ppm}$
for all seven resonance modes on all wafers, which speaks for the
reliability of the extraction method. It should be noted that without
correction of the angular misalignments (Section \ref{sub:Detection-of-angular}),
the least squares method would have resulted in errors up to 60~ppm.

Validity of the linearization step needed for the least square method
was confirmed: Frequency changes from approximation of (\ref{eq:linearized_01})
were calculated for each of the extracted elastic parameters $c_{ij}(T)$,
and compared with the non-linearized counterpart $\delta f_{k}^{\mbox{th}}(c_{ij}(T))$.
Linearization error was found to be less than 5 ppm for all cases.

Based on the above analysis, we expect that by using the extracted
elastic parameters, one can estimate the frequency of an arbitrary
resonance mode, fabricated on a wafer with similar carrier concentration
as in our experiments, with following accuracies: 
\begin{itemize}
\item Absolute frequency of a resonator can be predicted with $\pm1000\,\mbox{ppm}$
accuracy. 
\item Thermal drift over a temperature range of $T=-40\ldots+85^{\circ}\mbox{C}$
can be predicted with $\pm25\,\mbox{ppm}$ accuracy.
\begin{figure}[h]
\begin{centering}
\includegraphics[bb=20bp 80bp 650bp 500bp,clip,width=1\columnwidth]{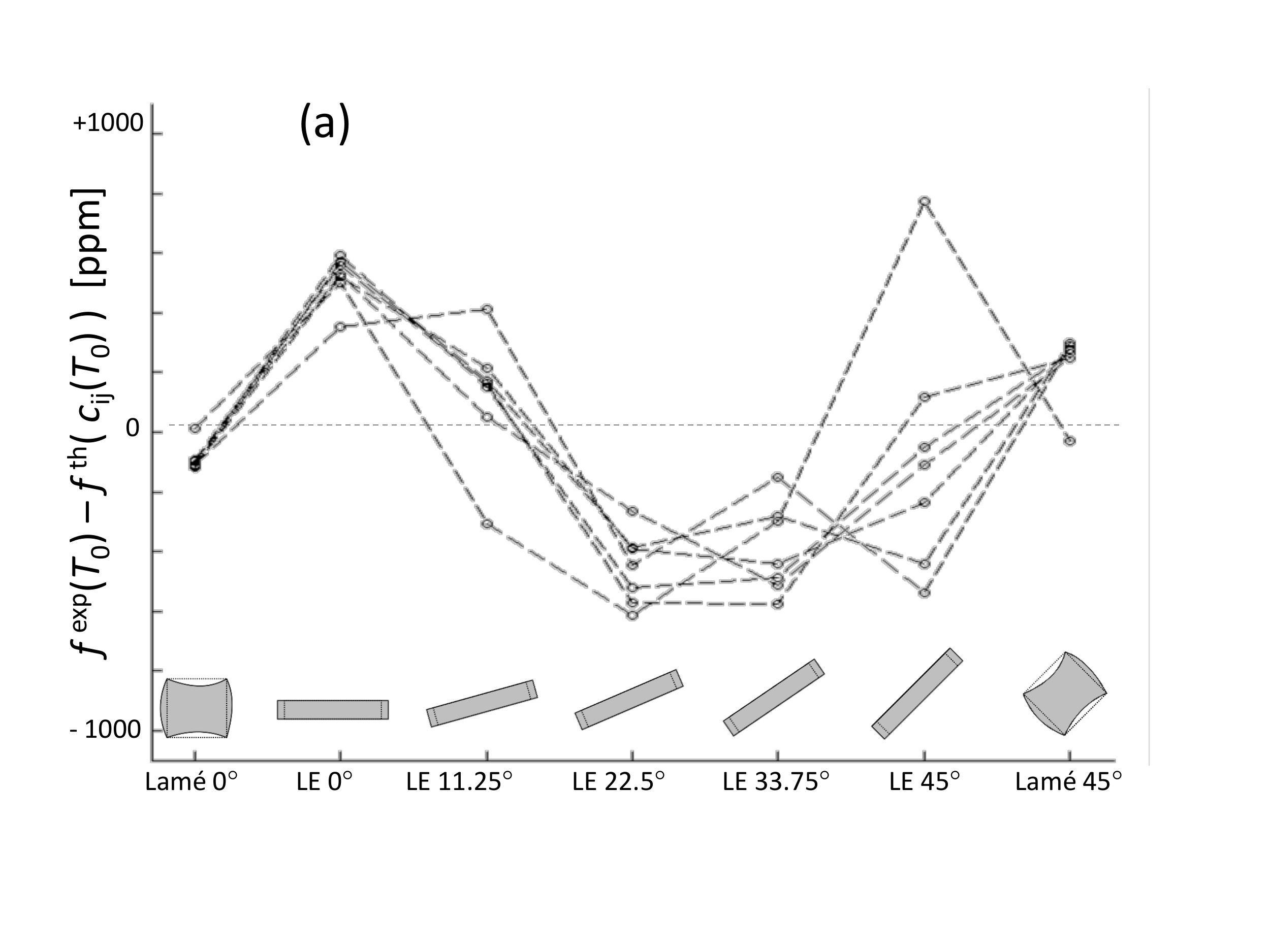}
\par\end{centering}

\begin{centering}
\includegraphics[bb=20bp 70bp 650bp 500bp,clip,width=1\columnwidth]{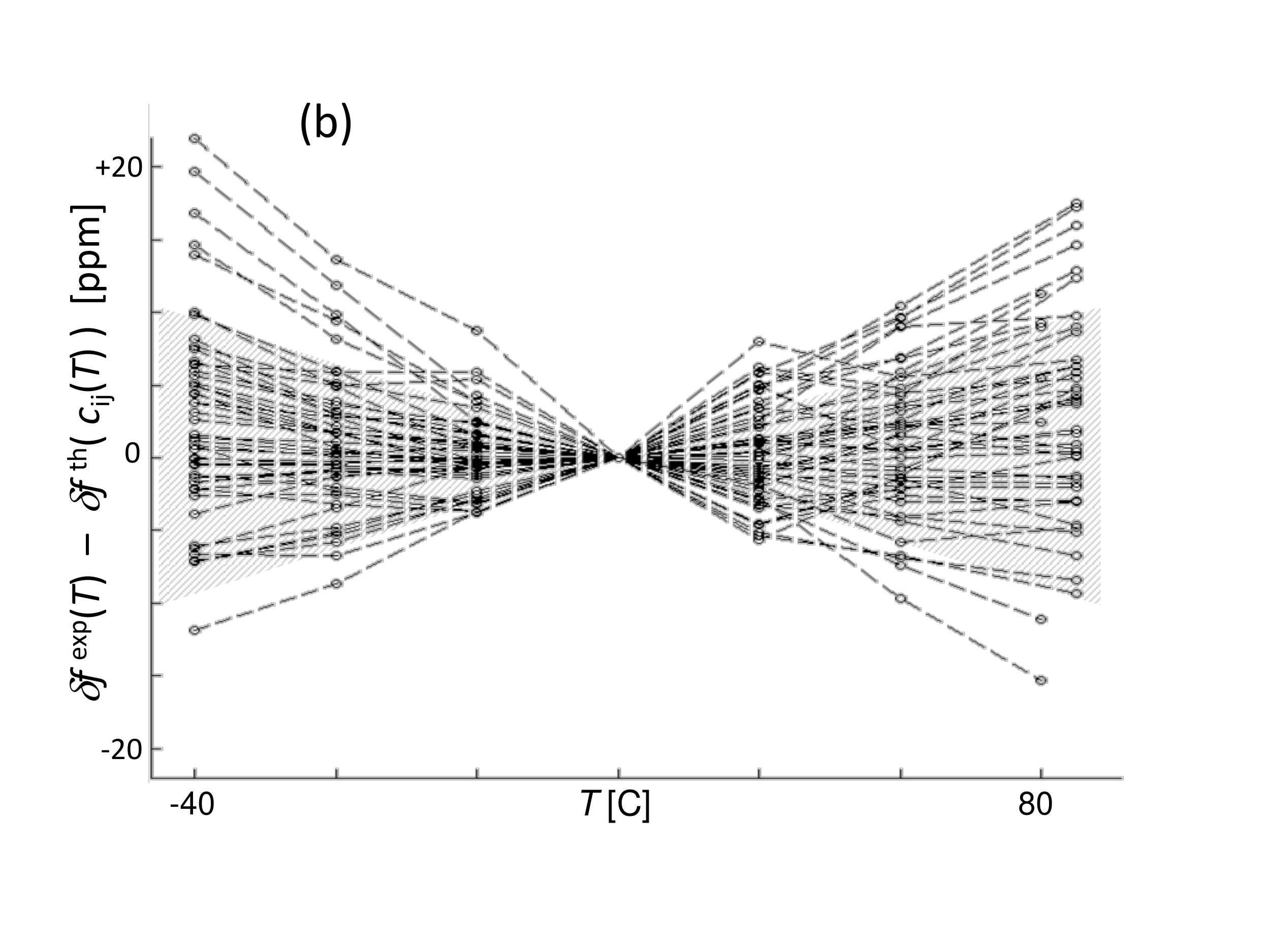}
\par\end{centering}

\caption{(a) Difference between measured frequencies $f_{k}^{\mbox{exp}}(T_{0})$
and theoretical estimates $f_{k}^{\mbox{th}}(c_{ij}(T_{0})$) at $T_{0}=20^{\circ}\mbox{C}$
for all resonance modes on all wafers. Theoretical estimates are based
on $c_{ij}(T_{0})$, which are the fitted elastic parameters obtained
through numerical minimization of (\ref{eq:optim_for_cij0}). (b)
Difference between the measured relative frequency changes $\delta f_{k}^{\mbox{exp}}(T)$
and the corresponding theoretical estimates $\delta f_{k}^{\mbox{th}}(c_{ij}(T_{0})+\delta c_{ij}(T))$
for all resonance modes on all wafers. Theoretical estimates are calculated
using $\delta c_{ij}(T)$ that have been fitted using (\ref{eq:LSQ}).
The shaded region illustrates the frequency measurement uncertainty
that has been assumed in error analysis of Section \ref{sub:Error-analysis}.
\label{fig:errrorfigs}}
\end{figure}

\end{itemize}

\subsection{Accuracy of temperature coefficients of $c_{ij}$ parameters\label{sub:Error-analysis}}

While the analysis of the previous section provides a way to establish
a confidence level on the resonance frequencies that can be calculated
from the extracted elastic parameters of this work, one can obtain
also estimates for the accuracy of the temperature coefficients of
elastic constants. Let us denote these confidence intervals as $\Delta c_{ij}^{0}$,
$\Delta a_{ij}$ and $\Delta b_{ij}.$ They are listed in Table \ref{tab:cij_results_table}
and also shown as vertical error bars in Fig. \ref{fig:CIJ_matrix}.

For $\Delta c_{ij}^{0}$, an upper limit of $\pm2000\,\mbox{ppm}$
is obtained by starting from the above discussed absolute frequency
accuracy of $\pm1000\,\mbox{ppm}$, and by applying (\ref{eq:frquency-general}).
Other potential error sources of smaller magnitude are:
\begin{itemize}
\item The mass of dopant atoms differs from that of silicon. Assuming that
the volume of the crystal stays constant the maximal density change
is less than 200~ppm, which would be reflected as a similar inaccuracy
in $c_{ij}^{0}$.
\item The dimensions of the resonators may vary slightly from the designed
measures due to potential mask bias and imperfections in DRIE etching.
A conservative estimate for the lateral dimension change of $\pm0.1\,\mbox{\ensuremath{\mu}m}$
of the resonators would deviate the resonance frequency approximately
by $\pm300$~ppm, and thus have an effect of $\pm600\,\mbox{ppm}$
on $c_{ij}^{0}$.
\item Thickness variation of $\pm2\,\mbox{\ensuremath{\mu}m}$ of the device
layer would have a very small effect on frequencies of the resonators:
FEM analysis indicated that the frequencies of the LE modes stay within
$\pm20\,\mbox{ppm}$, and changes are even smaller for the Lamé modes. 
\end{itemize}
To assess the inaccuracy of the first and second order temperature
coefficients of the elastic parameters, a Monte Carlo approach was
used to simulate the effect of several error sources. A large number
of copies of the experimental data sets (of Fig. \ref{fig:spiders})
were taken, and perturbed according to following sources of uncertainty,
labeled as $E_{1}\ldots E_{5}$:

$E_{1}$: Relative frequencies were deviated by $\delta f\times|\Delta T|/\Delta T_{\mbox{max}},$
where $\delta f$ was taken from a normally distributed population
with a standard deviation of 10~ppm, $\Delta T$ was defined as $T-20^{\circ}\mbox{C}$,
and $\Delta T_{\mbox{max}}$ was set to $65^{\circ}\mbox{C}$. The
distribution is visualized by the shaded region of Fig. \ref{fig:errrorfigs}(b)).
This way a distribution corresponding to the observed errors in relative
frequencies was reproduced.

$E_{2}$: The accuracy specified for the temperature controller and
chuck was taken into account by deviating the temperature points by
$\delta T$ taken from a normally distributed population with a standard
deviation of $0.5^{\circ}\mbox{C}$. 

$E_{3}$: Sensitivities of (\ref{eq:amatrix}) were evaluated at the
linearization points $c_{ij}^{0}$, which was estimated above to have
an uncertainty within $\pm2000\,\mbox{ppm}$. Error caused by this
was simulated by perturbing the linearization points accordingly.

$E_{4}$: Thermal expansion was assumed constant in the calculations,
and our measurements suggested this to hold for linear thermal expansion
within a $\pm7\%$ error marginal (\ref{sub:Doping-independency-of}).
The thermal expansion effect in (\ref{eq:thermalexp}) was perturbed
to take this uncertainty into account.

$E_{5}$: Thickness of the devices deviated from the nominal thicknesses
used in the simulations. Sensitivities of (\ref{eq:amatrix}) were
perturbed to take into account a thickness variation of $\pm2\,\mbox{\ensuremath{\mu}m}$.

Extraction of the elastic parameters $c_{ij}(T)$ was performed on
the perturbed data sets, identically to as it was done on the real
measurement data, and distributions of $a_{ij}$ and $b_{ij}$ were
obtained. Confidence intervals $\Delta a_{ij}$ and $\Delta b_{ij}$
were calculated for each error source $E_{1}\ldots E_{5}$ separately,
and the total effect was estimated as the rms sum. Values for $E_{1}$
and $E_{2}$ were obtained as standard deviations the $a_{ij}$/$b_{ij}$
distributions, while the full range was used for $E_{3}\ldots E_{5}$.
Inaccuracy of the first order coefficient $a_{ij}$ was found to range
from $\Delta a_{11-12}=\pm0.3\,\mbox{ppm/K}$ to $a_{12}=\pm1.5\,\mbox{ppm/K}$.
Correspondingly, error of the second order coefficient was seen to
vary from $\Delta b_{11-12}=\pm7\,\mbox{ppb/K}^{2}$ to $\Delta b_{12}=\pm33\,\mbox{ppb/K}^{2}.$
Error in measured frequencies ($E_{1}$) and inaccuracy of temperature
($E_{2}$) were major sources of uncertainty for all $a_{ij}/b_{ij}$
parameters, and the linearization point error ($E_{3}$) was a top
contributor for $a_{11}$ and $a_{12}$.

\subsection{Doping independency of thermal expansion\label{sub:Doping-independency-of}}

The procedure for extracting the elastic parameters relied on the
assumption that thermal expansion of (\ref{eq:thermalexp}) would
be insensitive to doping. To our knowledge, effects from heavy doping
to thermal expansion of silicon have not been studied experimentally.
For verification, mechanical dilatometry was used for measuring the
thermal expansion of samples with similar doping levels as wafers
B3, As1.7 and P7.5. The linear thermal expansion coefficient $\alpha_{1}$
was found to be constant within the $\pm7\%$ error marginal of the
measurement.

\subsection{Manufacturability of temperature compensated MEMS resonators}

\begin{figure}[h]
\begin{centering}
\includegraphics[bb=20bp 30bp 700bp 490bp,clip,width=1\columnwidth]{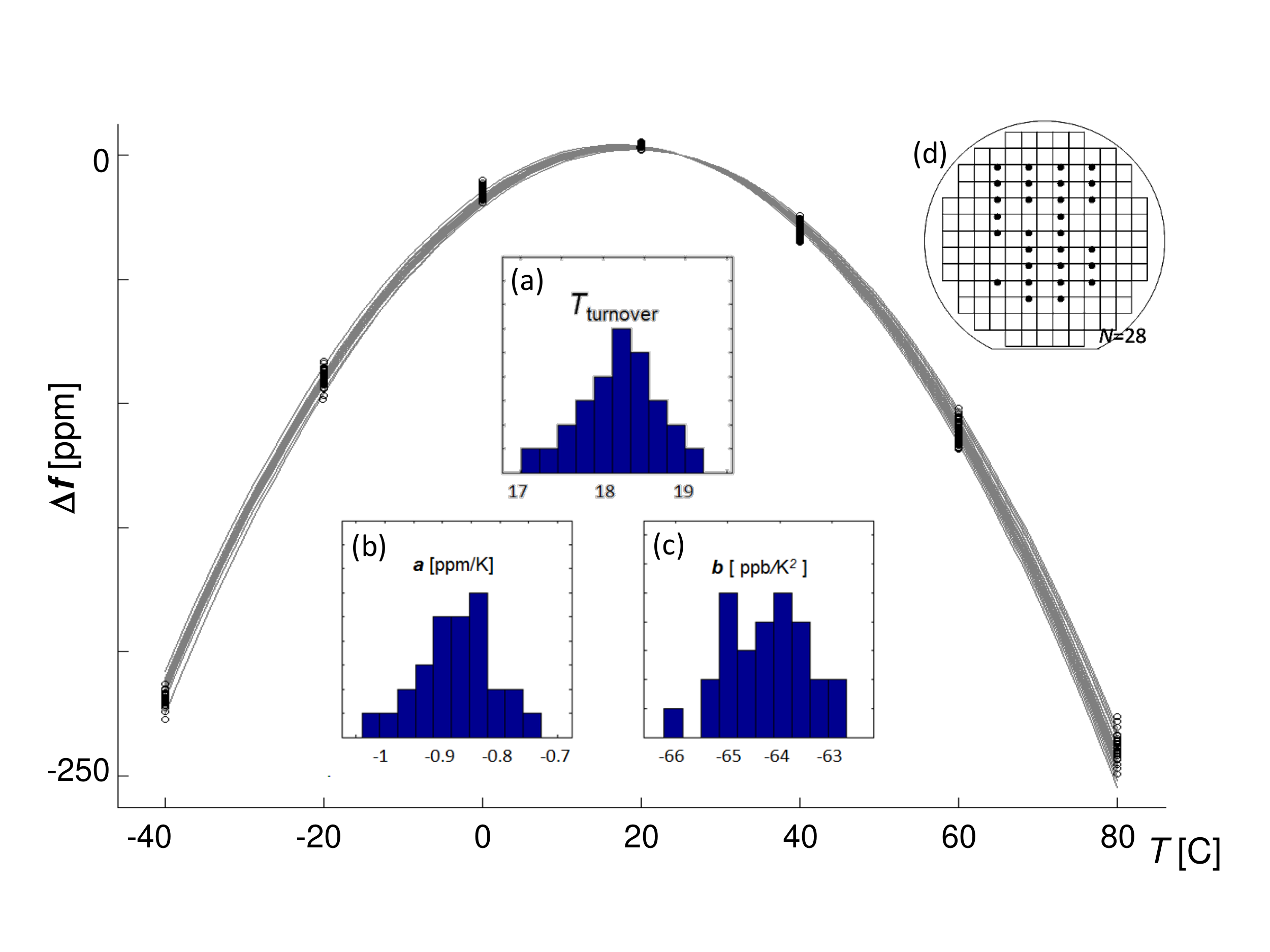}
\par\end{centering}

\caption{Superposed $f$ vs. $T$ curves of 28 square extensional resonator
samples similar to that of Ref. \cite{pensala_temperature_2011}.
Measurements are shown as black open circles, while second order polynomial
fits to the data are denoted with gray lines. The wafer has been doped
with phosphorus to a concentration of $n\sim5\times10^{19}\mbox{cm}^{-3}$.
Histograms (a), (b) and (c) illustrate the distribution of the turnover
temperature, and the first/second order temperature coefficients of
frequency, respectively. Wafer map (d) shows the location of the measured
devices on the wafer. \label{fig:stats_example}}
\end{figure}
Eventual manufacturability of silicon resonators whose temperature
compensation is based on degenerate doping crucially depends on the
statistical variations of the $f$ vs. $T$ curves among devices fabricated
on a single wafer or on a batch of wafers. This aspect was addressed
by studying a set of \textasciitilde{}30 square extensional mode resonators
of Ref. \cite{pensala_temperature_2011}, which were fabricated on
a wafer with a specification similar to that of wafer P4.7 of this
work. The $f$ vs. $T$ curves of the devices are shown in Fig. \ref{fig:stats_example}.
These devices were temperature compensated to first order with their
turnover temperatures near $20^{\circ}\mbox{C}$. The overall frequency
drift over the whole temperature range of $120\,\mbox{C}^{\circ}$
stays within 250~ppm, and maximum deviation between samples is approximately
20~ppm. One should note that the data of this example is from devices
on a Czochralski grown wafer, where the doping level may vary $\pm5\%$
within the wafer. A better control of doping level is achievable with
diffusion based doping, or with epitaxially grown silicon, where doping
can be controlled during the crystal growth process.

\section{Conclusion}

Elastic constants $c_{11}$, $c_{12}$ and $c_{44}$ of degenerately
doped silicon were studied experimentally as a function of the doping
level and temperature. First and second order temperature coefficients
of the elastic constants were extracted from measured resonance frequencies
of a set of MEMS resonators fabricated on wafers with varied doping.

The linear temperature coefficient of the shear elastic parameter
$c_{11}-c_{12}$ was found to be zero at n-type doping level of $n\sim2\times10^{19}\mbox{cm }^{-3}$.
It was observed to increase to over $+40\,\mbox{ppm/K}$ with higher
level of doping, which implies that the frequency of many types of
resonance modes, including extensional bulk modes and flexural modes,
can be temperature compensated to first order. The second order temperature
coefficient of $c_{11}-c_{12}$ was found to decrease by 40\% in magnitude
when n-type doping was increased from 4.1 to $7.5\times10^{19}\mbox{cm }^{-3}$,
suggesting a further reduction of the second order effect with increased
doping.

It was found that the frequency drift of an arbitrary silicon resonator
design, fabricated on a wafer with doping level similar to those investigated
in this work, can be estimated with an accuracy of $\pm25$~ppm over
a temperature range of $T=-40\ldots85^{\circ}\mbox{C}$ using the
elastic parameters of this work. Absolute frequency can be calculated
with an accuracy of $\pm1000$~ppm.

\section{Acknowledgments}

The authors would like to acknowledge the Finnish Funding Agency for
Technology and Innovation (Tekes), Okmetic Oyj, Murata Electronics
and Micro Analog Systems for funding. Okmetic Oyj is acknowledged
for providing the silicon wafers. A.J. acknowledges funding from the
Academy of Finland, and wishes to thank Arto Nurmela for help in the
measurements. Roger Morrell is thanked for performing the thermal
expansion measurements.

\bibliographystyle{IEEEtran}
\bibliography{Tcompensation_APL}

\end{document}